\documentclass{camera}
\def\be{\begin{equation}}
\def\ee{\end{equation}}
\usepackage{graphicx}

\begin{document}

\title{
SN1987A: Revisiting the Data and the Correlation between Neutrino and Gravitational Detectors}

\author{P.Galeotti$^a$, G.V.Pallottino$^{b}$ \& G.Pizzella$^{c}$}
 \organization{
 
{\it a\rm})Dipartimento di Fisica dell'Universit\`a di Torino \\INFN Sez. di Torino\\
 
{\it b\rm})  Dipartimento di Fisica dell'Universit\`a di Roma ``La Sapienza" \\ INFN Sez. di Roma\\
 {\it c\rm}) Dipartimento di Fisica, Universit\`a di Roma ``Tor Vergata''\\
 INFN Laboratori Nazionali di Frascati\\
}

\maketitle

\begin{abstract}
We re-examine the data taken by the neutrino detectors during the supernova SN1987A. It is found that the Kamiokande data, in addition to the well known burst at 7:35 hours UT, show another one at 7:54 hours, with seven pulses in 6.2 seconds. This second burst supports the idea that the duration of the collapse was much longer than a few seconds, as already suggested by the LSD detection at 2:56 hours the same day, i.e. four and a half hours earlier. The correlations between the gravitational wave detectors (Rome and Maryland) and the neutrino detectors are also revisited. It is shown that the g.w. detectors exhibit significant correlations with both the LSD and the Kamiokande detectors over periods of one-two hours that are centered, in both cases, at the LSD time.

\end{abstract}
\vspace{1.0cm}

\section{Introduction}

Supernova 1987A was a unique event during our time, since it occurred in LMC, i.e. relatively close to the Earth, when recently installed neutrino detectors were taking data: LSD in Italy, Kamiokande in Japan, IMB in the United States and Baksan in Russia.
The first neutrino burst was observed at 2:56 hours U.T. of 23 February 1987 by the detector LSD located in the Mont Blanc laboratory \cite{lsd}. This event, recorded on real time several hours before the optical detection of the supernova, was immediately reported on March 2 at the \it Rencontres de Physique de la Vall\`{e}e d'Aoste\rm . The SN visual observation then triggered the search for signals in the other neutrino detectors. A second neutrino burst was found at 7:35 hours of the same day in the data of Kamiokande \cite{ka1,ka2}, IMB \cite{imb} and Baksan \cite{baksan}.
The occurrence of two neutrino bursts, with time distance of about four and half hours, appeared surprising because the most accepted theories predicted that a star should collapse in a very short time, in the range of a few seconds or even less.
New theories were proposed suggesting that, because of the fragmentation of a fast rotating core, the phenomenon could have lasted for a few hours \cite{deru,casta}, thus allowing both the Mont Blanc and the Kamiokande neutrino detection events. More recently Imshennik and Ryazhskaya \cite{imm} have proposed the collapsar model, developing a detailed mechanism based on the idea that the collapsing star breaks under rotation in various pieces. In this way the emission of gravitational waves could occur for several hours, while the light fragments spiral around the collapsed massive central body.
In spite of these attempts to explain the experimental results, a large part of the scientific community persisted in the idea that the phenomenon should have lasted only a few seconds and the LSD observation was considered as due to chance. The problem however, remains open to investigation: does the collapse occur within a few seconds or it may last for some hours?
The latter alternative is now supported by new data analysis, since it has been
recently reported \cite{gapi} that the Kamiokande data show another burst, in addition to the well known one at 7:35 hours. This burst, discussed in the next section, consists of a sequence of seven pulses during 6.2 seconds, occurring at 7:54 hours, therefore in agreement with the hypothesis of a long duration of the phenomenon, as already suggested by the Mont Blanc observations and by the correlation with the data of the gravitational wave detectors.

\section{Neutrino bursts observed by the Kamiokande detector}

The data recorded by Kamiokande consist of a list of events characterized by their time of occurrence and by the parameter $N_{hit}$, which is the number of photomultipliers hitted, with a threshold set at $N_{hit}$ = 20 that roughly corresponds to a neutrino energy of 7.5 MeV.

We searched the event list provided to us by the Kamiokande collaboration for possible clusterings. We found two clusters, the first one being that reported by the Kamiokande group, of 11 pulses during 12.4 s starting at $7^h35^{min}33.7^{sec}$ U.T., with an extremely low imitation rate from the background. But we also found, unexpectedly, another cluster about 20 minutes later. This second cluster, as reported in Table 1, starts at $7^h54^{min}22.2^{sec}$ U.T. and consists of 7 pulses in a time window of 6.2 s with $N_{hit}$ ranging from 22 to 33, that is well above the threshold, and a background imitation rate of 1 event in 669 years.

 Since muons have been removed by the list of data we received from the Kamiokande collaboration, and since the possible effects of muons on the pulses constituting the first cluster have been studied very carefully by the Kamiokande group, we believe highly improbable that the second cluster of triggers (not discussed by the Kamiokande collaboration) be due to muons. We believe that this second burst escaped to the search of the Kamiokande team. As a matter of fact, one can find indication of it in the fig.4 of ref.3, from which, however, one does not realize, by looking to the figure, that it consists of seven pulses in just six seconds and well above background.
 
 In the Table \ref{tavek} we give the list of the pulses constituting this second burst.
  \begin{table}
\centering
\caption{The seven pulses of the Kamiokande second burst occurring within 6.2 s at 07:54 hours. Since the background corresponds to 0.024 pulses per second above $N_{hit}$ = 20, the probability to have such a cluster is once in 669 years.}
\vskip 0.1 in
\begin{tabular}{|c|c|}
\hline
\hline
hour~min~sec&$N_{hit}$\\
&\\
\hline
\hline    
    7~  54~ 22.26&   33\\
    7~  54~ 24.11&   29\\
    7~  54~ 25.33&   28\\
    7~  54~ 25.34&   27\\
    7~  54~ 27.13&   22\\
    7~  54~ 28.37&   22\\
    7~  54~ 28.46&   22\\
 \hline
 \hline
\end{tabular}
\label{tavek}
\end{table}

We recall that the IMB detector, whose energy threshold was above 20 MeV, gave signals in coincidence with the first Kamiokande burst, but not reported coincidences at other times. This is probably due to the fact that the \it canonical \rm Kamiokande burst at 7:35 hours consisted of several high energy signals, well above that IMB threshold, while the  second burst has pulses below $\sim$ 15 MeV, i.e. below the IMB threshold.

\section{Correlation of LSD and Kamiokande with the gravitational wave detectors}

At the time of the SN1987A the cryogenic resonant gravitational wave detectors were not ready yet, still in the construction phase. However in Rome the room-temperature resonant detector GEOGRAV, intended to detect signals correlated with the Earth movements, was in operation. 
The Rome group was informed immediately by Carlo Castagnoli that the LSD neutrino detector had observed a cluster of five neutrino signals, with very low Poissonian probability to be accidental, at $2^h56^{min}36^{sec}$ U.T. of 23 February 1987. On the next day, since GEOGRAV was in operation in the best possible noise condition, although this detector was not sensitive enough for a possible g.w. according  to classical estimation of the cross-section, we carefully studied the data and found a correlation with the five-neutrino burst, with the g.w. signals anticipating the neutrino signals by 1.4 seconds. This result  was presented at the La Thuile meeting \cite{geo} on 3 March 1987.

On 7 March we learned about the Kamiokande observation of a large neutrino cluster occurring about four and half hours after the Mont Blanc neutrino burst. In spite of the difficulty due to the Kamiokande observation at a later time, coincident with observation made with the IMB experiment, we thought important to continue the study of the GEOGRAV data, since there was a great chance that no other visible Supernova would have occurred for the next hundred years or so. In addition, also Joe Weber ($\sim$6000 km away) had made observations with his room temperature detectors, and these appeared to have some degree of correlation with GEOGRAV.

The key idea for our analysis was to consider all the triggers recorded by the neutrino  detectors\footnote{In the following, for simplicity, we use the word $neutrino$ to indicate a pulse from the neutrino detectors, being aware that in most cases these triggers are due to background.}, that is including those usually discarded as noise when not grouped together. This allowed us to analyze all the available data, and not just those occurring near the time of the Mont Blanc burst at $2^h56^{min}36^{sec}$ U.T. To do this we used the following correlation algorithm \cite{somma} based on summing the energies of the two g.w. detectors at the occurrence times of the neutrino triggers, taking into account a possible common time shift between g.w. data and neutrinos. We calculate:
\be
E(\phi)=
\frac{1}{N_\nu} \sum_i^{1,N_\nu}[E_R(t_i+\phi)
+E_M(t_i+\phi)]
\label{frasca}
\ee
with the following meaning for symbols:\\
$N_\nu$ is the number of considered neutrino triggers in a given time period for the analysis (say one
hour),\\
$t_i$ indicates the time of the $i^{th}$ neutrino trigger,\\
$E_R$ and $E_M$, expressed in kelvin, are the measured energy innovations , obtained with a
Wiener-Kolmogoroff filter, from the data of the Rome (R) and the Maryland (M) g.w. detectors at the
times $t_i+\phi$ within $\pm0.5~s$,\\
$\phi$ is a time shift common to the two g.w. detectors.

The statistical significance of the results obtained with the above algorithm is checked by comparing, for each value of $\phi$, the value of $E(\phi)$ with the M values $E(random)$ obtained from the same equation (1) by adding random time shifts $\phi_1$ and $\phi_2$ separately, to the two g.w. data streams, representing therefore cases with no match between the times of the g.w. measurements and the occurrence times of the neutrino triggers. If a correlation exists at a common time delay $\phi$ we expect the value $E(\phi)$ be the largest, or one of the largest, among the M random values $E(random)$ used as reference background. We remark that M can be made very large, thanks to the addition of the independent shifts $\phi_1$  and $\phi_2$  to the data of the two g.w. detectors, thereby providing a statistically rich reference.

The results of the analysis show a very strong correlation between the g. w. detectors and the LSD neutrino detector, with an optimum time shift  $\phi_o\sim -1.2~s$ (the g.w. signals preceding the neutrino triggers), lasting for a period of about two hours centered at the LSD time. The time shift of $\phi_o\sim -1.2~s$, determined here with 97 neutrino triggers, was only 0.2 seconds off from our result presented several months before, when we used only the five-neutrino burst. This result was presented at the La Thuile meeting in March 1988, and published on Il Nuovo Cimento \cite{og1}.

\begin{figure}
\includegraphics[width=0.8\linewidth]{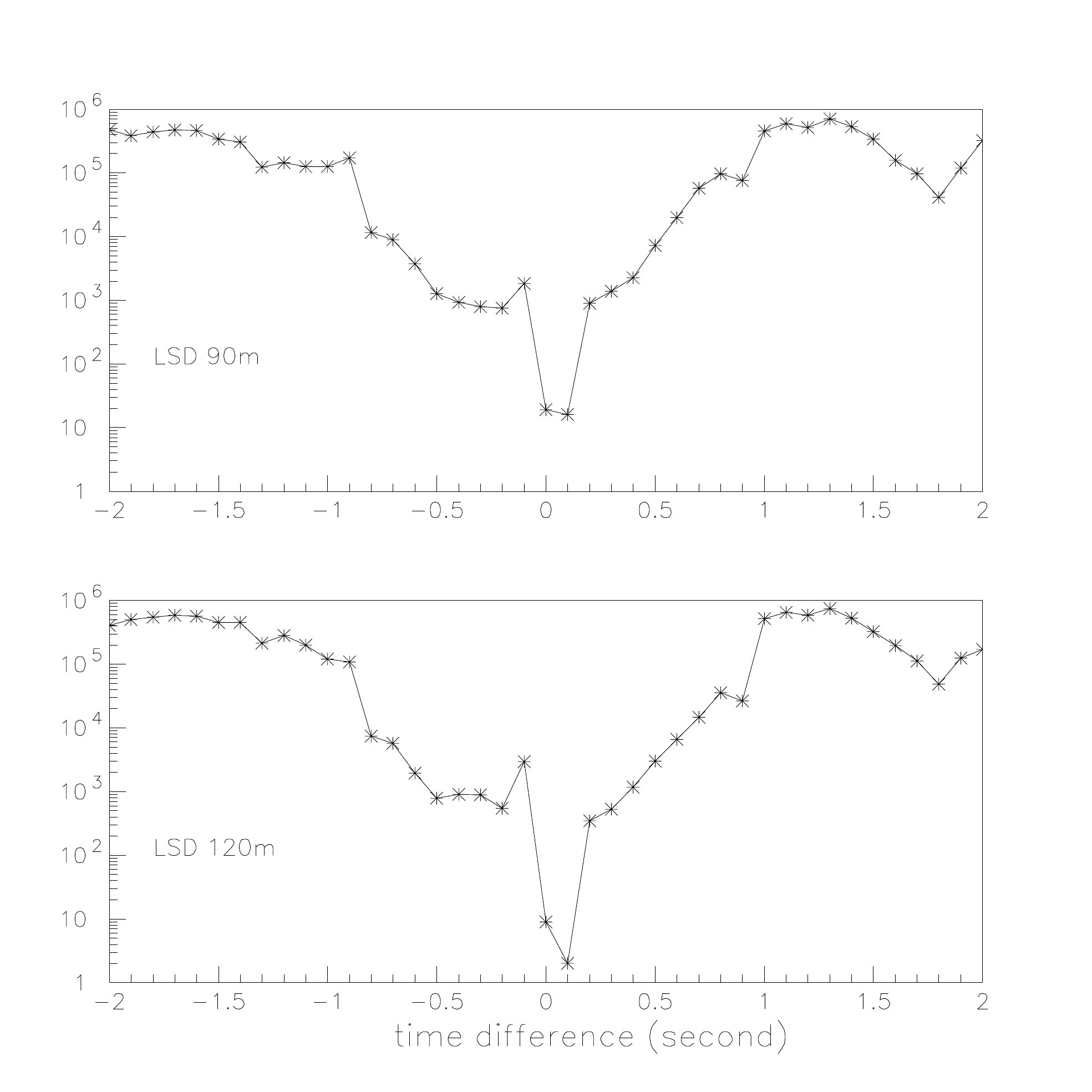}
 \caption{Correlation between the  gravitational wave detectors with the Mont Blanc neutrino detectors for two different time periods (90 minutes and 120 minutes) both centered at the LSD time. The plots report the number of random trials (out of one million) giving $E(random)\ge E(\phi)$, versus the time shift $\phi+1.2~s$.
        \label{correlamb} }
\end{figure}

In fig.\ref{correlamb} we report the results for two periods of analysis, one and half hour and two hours. On the abscissa we have the quantity $\phi+1.2~s$. On the ordinate there is the number of random trials (out of one million) giving $E(random)\ge E(\phi)$. The maximum correlation for the two-hour period occurs at  $\phi+1.2=0.1~s$, that is for $\phi_o=-1.1~s$. We have used $10^6$ random determinations of the background, thus $\frac{2}{10^6}$ is an estimate of the probability that the correlation is due to chance.

   \begin{figure}
\includegraphics[width=0.8\linewidth]{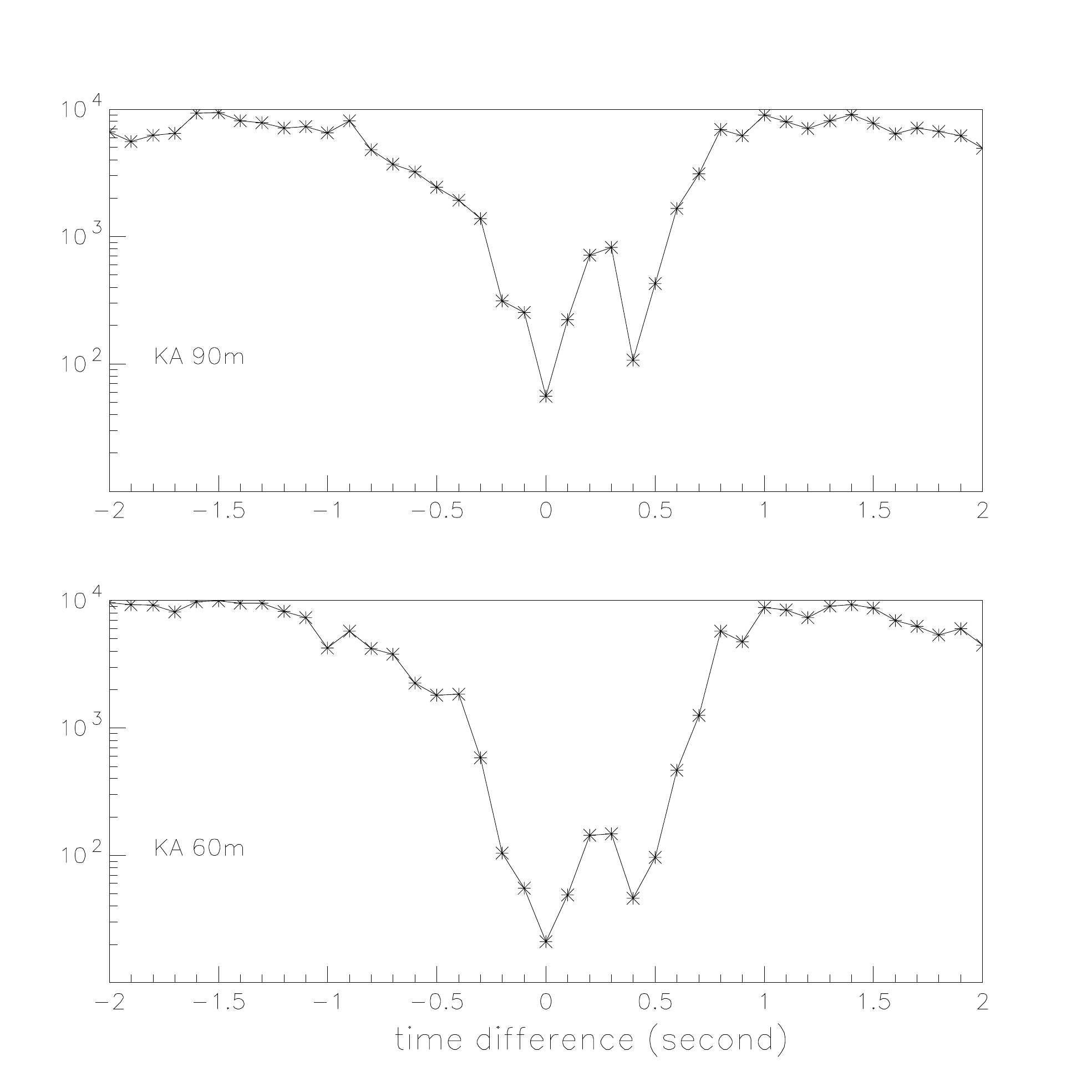}
 \caption{Correlation between the gravitational wave detectors and the Kamiokande neutrino detector (with a time correction of +7.8 s) for two different periods (60 minutes and 90 minutes) both centered at the LSD time. The plot reports the number of random trials (out of ten thousand) giving $E(random)\ge E(\phi)$, versus the delay $\phi$ (shifted by 1.2 s)
        \label{correlaka} }
\end{figure}

   \begin{figure}
\includegraphics[width=0.8\linewidth]{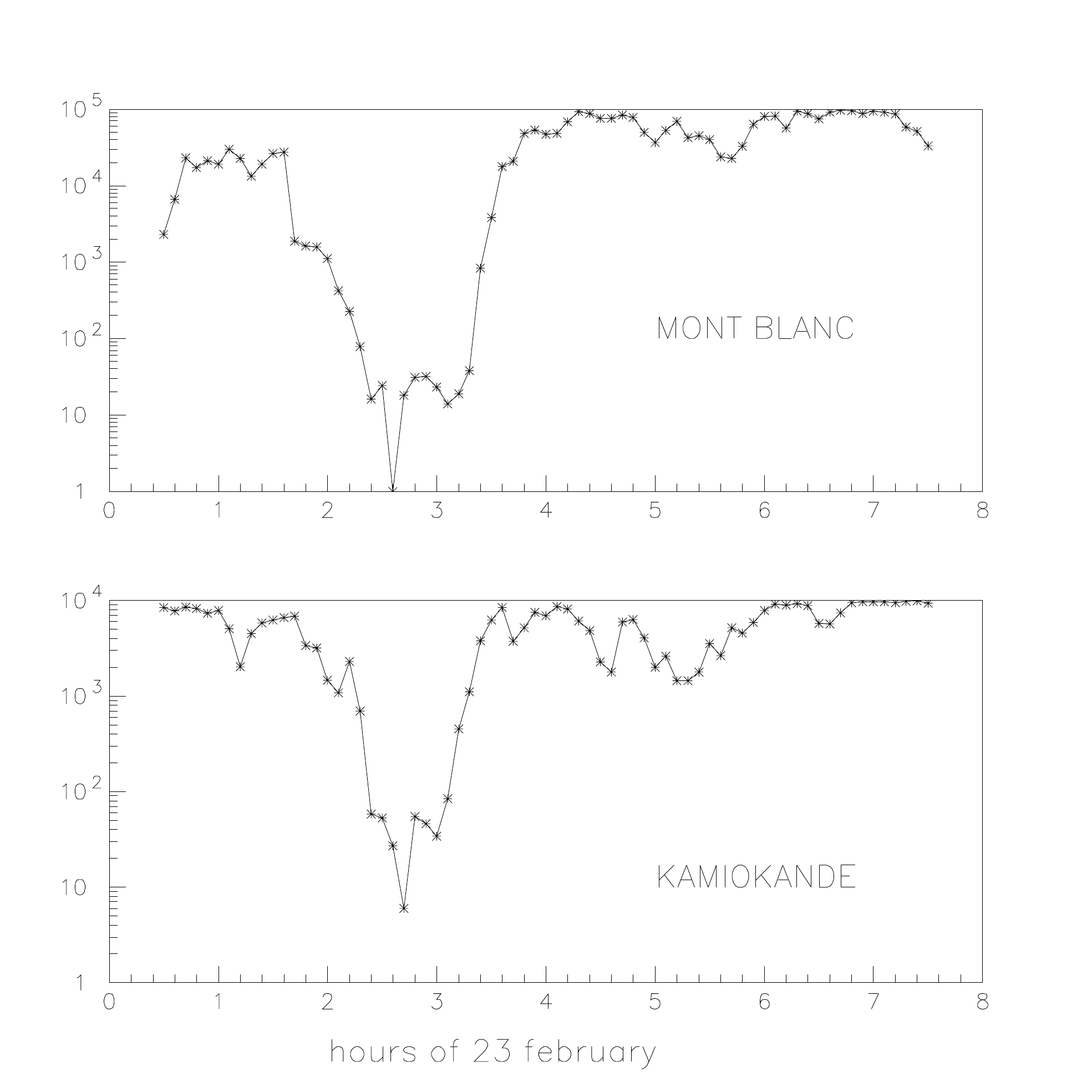}
 \caption{Correlations between the g.w. detectors and the LSD and Kamiokande neutrino detectors from 0.5 to 7.5 UT hours during one-hour running time periods. 7.8 seconds have been added to the recorded Kamiokande time. The plots report the number of random trials (out of $M=10^5$ for LSD and $M=10^4$ for Kamiokande) giving $E(random)\ge E(\phi)$, for a fixed value of $\phi_o= -1.2~s$.
        \label{corre1ora} }
\end{figure}

At this point we thought important to apply the same algorithm to the Kamiokande data. These data, recorded for an experiment aimed at the measurement of the proton lifetime, had a time uncertainty of $\pm$1 minute but the time could be adjusted by imposing a coincidence with the IMB event at 7:35 hours. This correction was 7.8 s \cite{sch}.
We obtained from Kamiokande a magnetic tape with the complete list of 31365 events (covering one full day) and found a correlation with Kamiokande provided we added 7.8 s to the Kamiokande originally recorded time.

In fig.\ref{correlaka} we present the results, see ref.\cite{og2}, of this analysis for two periods: one and half-hour and one hour, both centered at the LSD time. On the abscissa we have the quantity $\phi+1.2~s$, thus the maximum correlation for the one-hour plot occurs at $\phi+1.2=0.0~s$ that is for $\phi_o=-1.2~s$. Here we have used $10^4$ random determinations of the background, thus $\frac{2}{10^4}$ is an estimate of the probability that the correlation is due to chance. We note that the correlation for the Kamiokande data is weaker than for LSD and extends over shorter time periods.

More recently we have considered a possible correlation in the entire period under study, that is from 0:00 UT hour and 8:00 UT hour. The correlation is calculated over one-hour time intervals, running from 0.5 to 8 U.T. hours of February $23^{rd}$ in steps of 0.1 hour, and is shown in fig.\ref{corre1ora}. The time of the Kamiokande experiment has been adjusted by 7.8 second. 
 
Both neutrino detectors show a clear correlation with the g.w. detectors
at the same time, the Mont Blanc event time, with a striking similarity
between them. In addition, it is important to notice that the Kamiokande
correlation with the g.w. detectors at the Mont Blanc time occurs without
using any data of the Mont Blanc experiment, but only the time correction
from IMB.

\section{Discussion and conclusion}

We have shown experimental evidence that the phenomenon connected with the SN1987A has a duration of the order of a few hours. This statement is supported by the following facts:
\begin{enumerate}
\item
 the observation by LSD (threshold of 5 MeV) of neutrino signals about four hours and half before the occurrence of the neutrino signals observed by Kamiokande (threshold of  7.5 MeV), IMB (threshold of 15 MeV)  and Baksan (threshold of 10 MeV) detectors;
\item
 the observation of at least two very different significant bursts by the Kamiokande apparatus at a time distance between them of about 20 minutes;
\item
 the correlation observed between the g.w. detectors and the neutrino detectors, \bf both \rm  LSD and Kamiokande, at the \bf same \rm LSD time ($2^h56^{min}36^{sec}$ U.T.).
\end{enumerate}

The measurements by LSD at an early time can be explained by the lower threshold  and by the use of iron in the LSD detector (see discussion in ref.\cite{imm}).
It remains the problem for the g.w. observation, which lies in the small cross-section, according to the classical theory\footnote{Attempts have been made to consider a cross-section derived by different physical models, see ref.\cite{prepa,mole,yoghi}}. With this cross-section the signals observed with the g.w. detectors require a total conversion into g.w. of at least one thousand solar masses, which is impossible, provided we exclude that the g.w. have been produced in a beam and we were lucky to intercept \cite{yuri}.

We remark that the correlations reported here appear as due to a common cause acting on both the g.w. and the neutrino detectors, so that we cannot exclude the effect of other causes than g.w. (exotic particles ?).

\section{Acknowledgment}
We thank the Kamiokande collaboration for having supplied their data and Oscar Saavedra for useful discussions and suggestions.

\end{document}